\documentstyle[psfig]{mn}

\ifnfsstwo

\fi

\ifnfssone
  \newmathalphabet{\mathit}
    \addtoversion{normal}{\mathit}{cmr}{m}{it}
    \addtoversion{bold}{\mathit}{cmr}{bx}{it}

\fi

\ifoldfss

\fi

  \makeatletter
  \ifx\CUP@mtlplain@loaded\undefined  
  \makeatother  
    
  \else
    
  \fi

\loadboldmathitalic 
\loadboldgreek      
 
\def\etal{{\rm et al.~}}
\def\simless{\mathbin{\lower 3pt\hbox
   {$\rlap{\raise 5pt\hbox{$\char'074$}}\mathchar"7218$}}}   
\def\simgreat{\mathbin{\lower 3pt\hbox
   {$\rlap{\raise 5pt\hbox{$\char'076$}}\mathchar"7218$}}}   
\def\etal{{\rm et al.}}

\def \pc3 {pc$^{-3}$}

\makeatletter
\ifx\CUP@mtlplain@loaded\undefined  
  \newfont\bit{cmbxti10 at 9pt}
\else
  \newfont\bit{mtbxti10 at 9pt}
\fi
\makeatother  
 
\title[Free floating planets] {Free-floating planets in stellar clusters?} 
\author[Smith \& Bonnell]
{Kester W. Smith$^{1,2}$ \& Ian A.Bonnell$^{3}$ \\ 
$^{1}$ Institute of Astronomy, ETHZ, Z\"urich CH-8092, Switzerland. \\ 
$^{2}$ Paul Scherrer Institut, CH-5232 PSI-Villigen, Switzerland. \\
$^{3}$ School of Physics and Astronomy, University of St Andrews,
North Haugh, St Andrews, Fife, KY16 9SS, Scotland.
}
\date{Received 1 December 2000, Accepted }
\pagerange{\pageref{firstpage}--\pageref{lastpage}}
\pubyear{    }

\def\LaTeX{L\kern-.36em\raise.3ex\hbox{a}\kern-.15em
    T\kern-.1667em\lower.7ex\hbox{E}\kern-.125emX}

\begin{document}

\label{firstpage}

\maketitle

\begin{abstract}

We have simulated encounters between planetary systems and single
stars in various clustered environments. This allows us to estimate
the fraction of systems liberated, the velocity distribution of the
liberated planets, and the separation and eccentricity distributions
of the surviving bound systems. Our results indicate that, for an
initial distribution of orbits that is flat in log space and extends
out to 50AU, 50\% of the available planets can be liberated in a
globular cluster, 25\% in an open cluster, and less than 10\% in a
young cluster. These fractions are reduced to 25\%, 12\% and 2\% if
the initial population extends only to 20AU.  Furthermore, these
free-floating planets can be retained for longer than a crossing time
only in a massive globular cluster. It is therefore difficult to see
how planets, which by definition form in a disc around a young star,
could be subsequently liberated to form a significant population of
free floating substellar objects in a cluster.

\end{abstract}

\begin{keywords}
stars: formation -- stars: dynamics -- planets

\end{keywords}

\section{Introduction}

The discovery of numerous extrasolar planetary systems in the solar
neighbourhood (Mayor \& Queloz~1995, Marcy \& Butler~1996; Marcy~1999)
has revolutionised our ideas of the planet formation process and how
it can vary from system to system. Specifically, the fact that most of
the systems found contain relatively massive planets at small
separations, in contrast to our solar system, has engendered significant
research into possible orbital migration (e.g. Lin, Bodenheimer and
Richardson~1996). More recently, the discovery that there appears to
be no such close systems in the globular cluster 47 Tuc implies a
significant difference in planetary formation which could be due to
the stellar environment (Brown et al 2000, Gilliland et al 2000). Indeed, it is possible
that stellar interactions in the early stages of the globular cluster
were able to disrupt the circumstellar discs before any planets were
able to form (Bonnell \etal~2000) or that the increased radiation from
the expected number of O stars was sufficient to remove these
circumstellar discs before any planets could form (Armitage~2000).
Encounters with passing stars in a dense stellar environment can lead
to disruption of the planetary system and thus the ejection of the
planets (see e.g. Sigurdsson, 1992). 
This could lead to a population of free-floating planets in the cluster.
Recently there has been a reported detection of a population of
substellar objects in $\sigma$ Orionis (Zapatero-Osorio et al, 2000)
that could be due to stellar encounters. In this letter, we
investigate the formation of a population of free-floating 
planets in various cluster environments. We pay particular attention to the 
velocity distribution of this population, and the question 
of whether the bulk of the liberated objects could be retained 
in their natal environment once they are ejected from their parent system.

In the next section we discuss the properties of the initial planet
population and of the various clusters.  We then briefly summarise the
issue of interaction cross sections, including discussion of the
different possibilities following an interaction.  We then describe
the simulations of the various encounters and derive
velocity dispersions and other properties for both the free floating
and bound planet populations.

\section{Initial conditions}

Observations indicate that YSO disks are typically 100AU in radius
(McCaughrean \& O'Dell, 1996). Although it isn't clear to what radius
in the disk planets generally form, we can estimate based on our own
solar system that planet and planetessimal formation has occured at
radii out to 40-50 AU. In contrast, the extrasolar planets found so
far have been in orbits as tight as 4 days. These observations provide
us with the plausible range of planetary orbits to investigate.  The
inner end of this range is unlikely to be strongly affected by
encounters (Bonnell et al 2000), although it is possible that in
sufficiently dense systems, stellar encounters are able to disrupt the
planetary discs before the planets have formed or before they are able to
migrate to these small separations.  Furthermore, in the case of a
young cluster the close orbits may not yet be populated as the
migration timescale is of order 10$^7$ years or more (Lin, Bodenheimer
\& Richardson, 1996). We therefore consider planetary orbits between 1
and 50 AU in radius. The initial orbits are all circular, and the
distribution of separations is flat in log space. 
To restrict the parameter space studied, the 
parent and perturbing stars are assumed be of equal mass, either 0.7
or 1.5M$_{\odot}$. These masses were chosen as representative of 
the bulk of the expected stellar population.  

We consider three different cluster environments. The properties of
these are summarized in Table~\ref{clusters}. Our cluster environments
are intended to correspond to a globular cluster (dense and
long-lived, with a high velocity dispersion), an open cluster
(more diffuse with a much lower velocity dispersion), and a
young cluster (intended to correspond to conditions in dense star
forming regions such as the Trapezium. - see e.g. Clarke, Bonnell \&
Hillenbrand, 2000). The impact parameters were drawn from the expected
probability distribution for the cluster environment. This is calculated 
using the mean time between encounters given by Binney
and Tremaine (1987),
\begin{equation}
\frac{1}{t_{{\rm enc}}} = 16 \sqrt{\pi} nv_{{\rm disp}} R^{2}_{{\rm enc}} 
 \left(1 + \frac{GM_{*}}{2v^{2}_{{\rm disp}} R_{{\rm enc}}} \right).
\label{probeqn}
\end{equation}
Here, $t_{{\rm enc}}$ is the mean time between encounters within a
distance $R_{{\rm enc}}$, $n$ is the number density of stars in the
cluster, and $v_{{\rm disp}}$ is the velocity dispersion.

\begin{table}
\begin{center}
\begin{tabular}{c|c|c|c|c|c} \hline
Cluster  &Density     & V$_{disp}$        & Lifetime  & \multicolumn{2}{c}{$b$ (AU)}\\
         & pc$^{-3}$        & km~s$^{-1}$ &  years    &   Min          &       Max  \\
Globular & 10$^3$           &  10         &  10$^{9}$ &    3.43         &  24.26       \\
Open     & 10$^2$           &   1         &  10$^{9}$ &    33.32        &  221.22         \\
Young    & 5$\times$10$^3$  &   2         &  5$\times$10$^6$ & 47.27     & 328.09      \\ \hline
\end{tabular}
\caption{\label{clusters} The properties of the types of clusters
studied. The minimum and maximum impact parameters, $b$, are also shown.
These correspond to roughly 10\% and 99\% encounter probabilities for
each cluster. }
\end{center}
\end{table}

\section{Simulations}

Simulations of restricted three body motion were carried out using a
4th order Runge-Kutta code with adaptive stepsize on the ETH's {\em
Asgard} cluster \footnote{ Asgard is an Intel Pentium III {\em
Beowulf} cluster located at the ETH in Z\"urich.  It comprises 502
CPUs on 251 Dual-CPU nodes. }. The Runge-Kutta code was found to
conserve energy over the interactions to a few parts in 10$^5$ or
better. Initial planetary orbits were selected at random from the
log-flat distribution. The planetary orbits were isotropically
distributed with respect to the stellar orbit.  The stellar orbit was
started at a point where the potential energy of the stellar system
was 1\% or less of the kinetic energy. The planets were injected when
the force due to the perturber reached 1\% of the force due to the
parent. The simulations ran for a fixed time (usually 24 hours).
Closest approach typically occurred at a time between 10\% and 40\%
through the total simulation.  A total of 7860 systems were used for
each cluster.

\section{Results}

Table~\ref{results} shows the number of planets which became unbound,
remained bound to the parent star, or were exchanged during the
simulation.  As expected, a substantial number of planets were unbound
in the dense, long-lived globular cluster environment, fewer in the
open cluster case, and less than 10\% in the young cluster case.  More
disruption occurred for the high-mass stars than the lower mass case.

\begin{table}
\begin{center}
\begin{tabular}{c|c|c|c|c|c} \hline
Cluster             & \multicolumn{3}{c}{Ionised} & Survived  & Exchanged \\
                    &  Total &   Kept   & Lost           &           &          \\
Globular            &   47.3\% & 30.1\% & 17.2\%         & 51.5\%    & 1.3\%   \\     
Open                &   26.6\% & 0.5\%  & 26.1\%         & 61.1\%    & 12.3\%   \\
Young               &   7.8\%  & 0.5\%  &  7.3\%         & 90.1\%    & 2.1\%   \\ \hline
\end{tabular}
\caption{\label{results} 
The fate of planets in different cluster environments. In the case
of ionisation, three fractions are shown.
The total percentage of systems ionised, the percentage which
are retained in the cluster, and the percentage which escape
within a crossing time.  }
\end{center}
\end{table}

\subsection{Velocity distributions of free floating planets}

For the planets that became unbound, the velocity at infinity was
estimated from the total energy (kinetic plus potential) of the planet
and, assuming this energy is conserved while the planet escapes from
the gravitational potential, then $1/2 m v^2 = E_{\rm tot} \ge 0$.
The simulations with high-mass stars produced more high-velocity
liberated objects, but the difference in the final velocity
distributions was not large.

\begin{figure}
\vbox{
\psfig{{figure=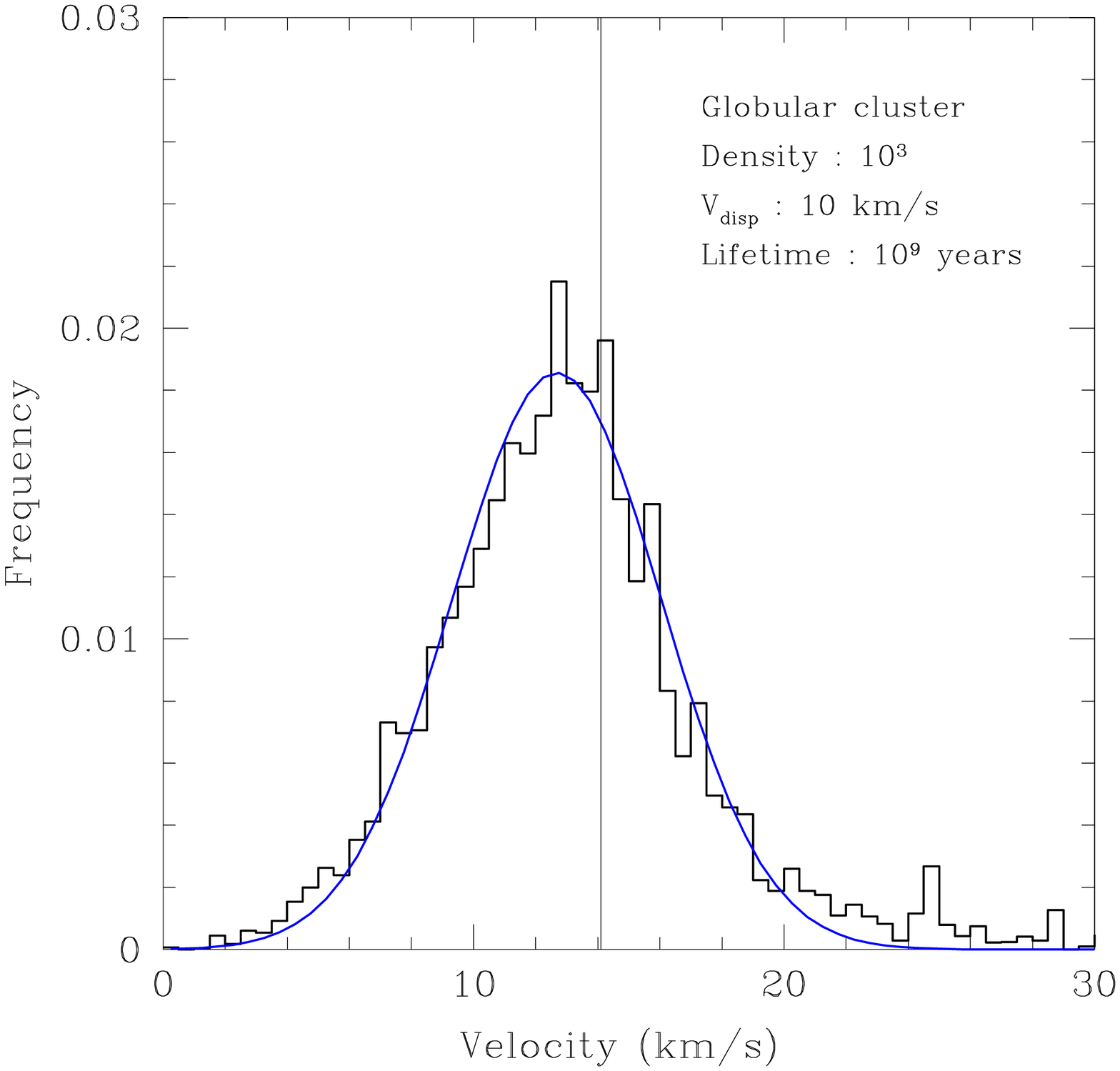,width=3.truein,height=2.3truein}}
\psfig{{figure=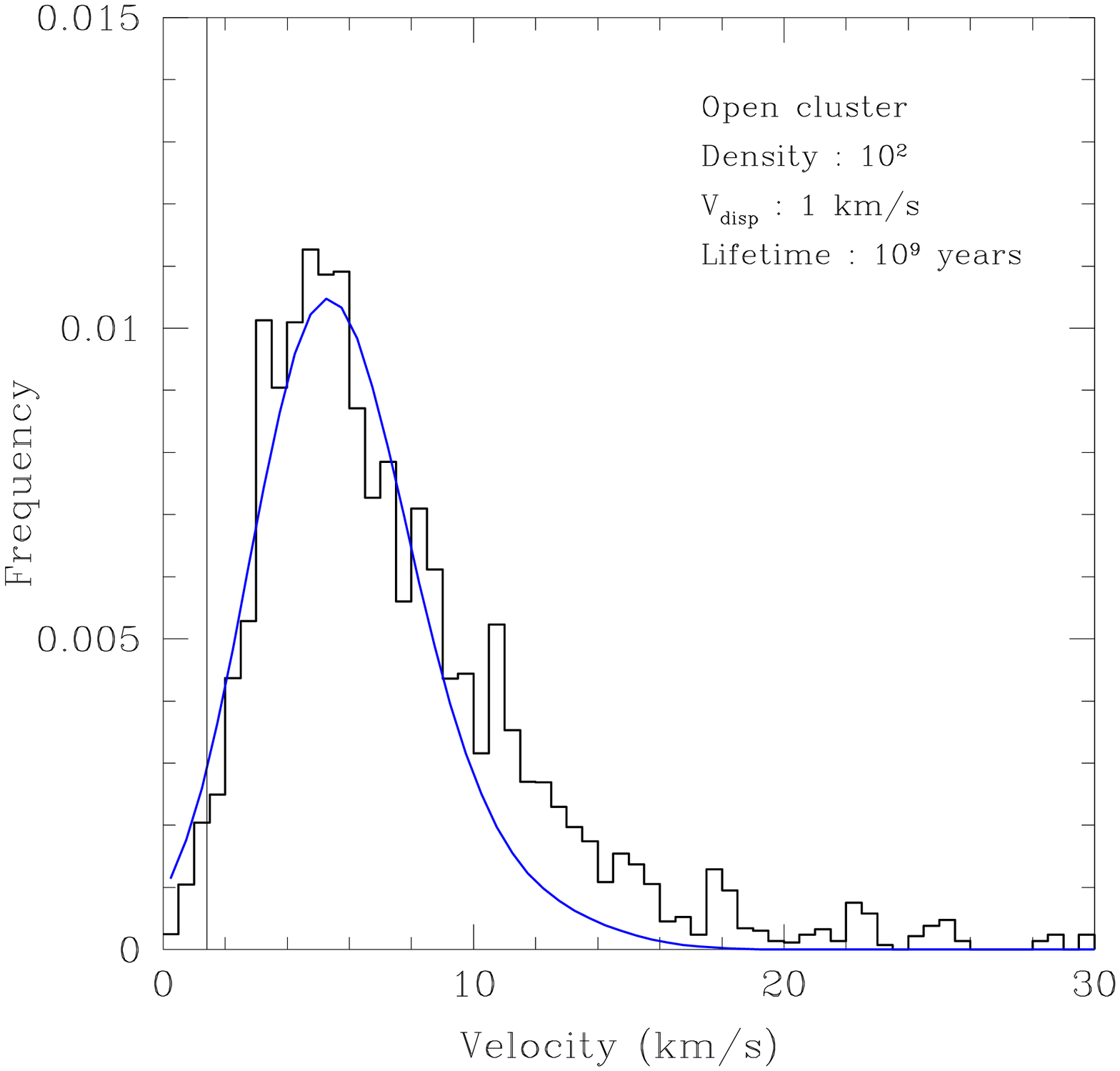,width=3.truein,height=2.3truein}}
\psfig{{figure=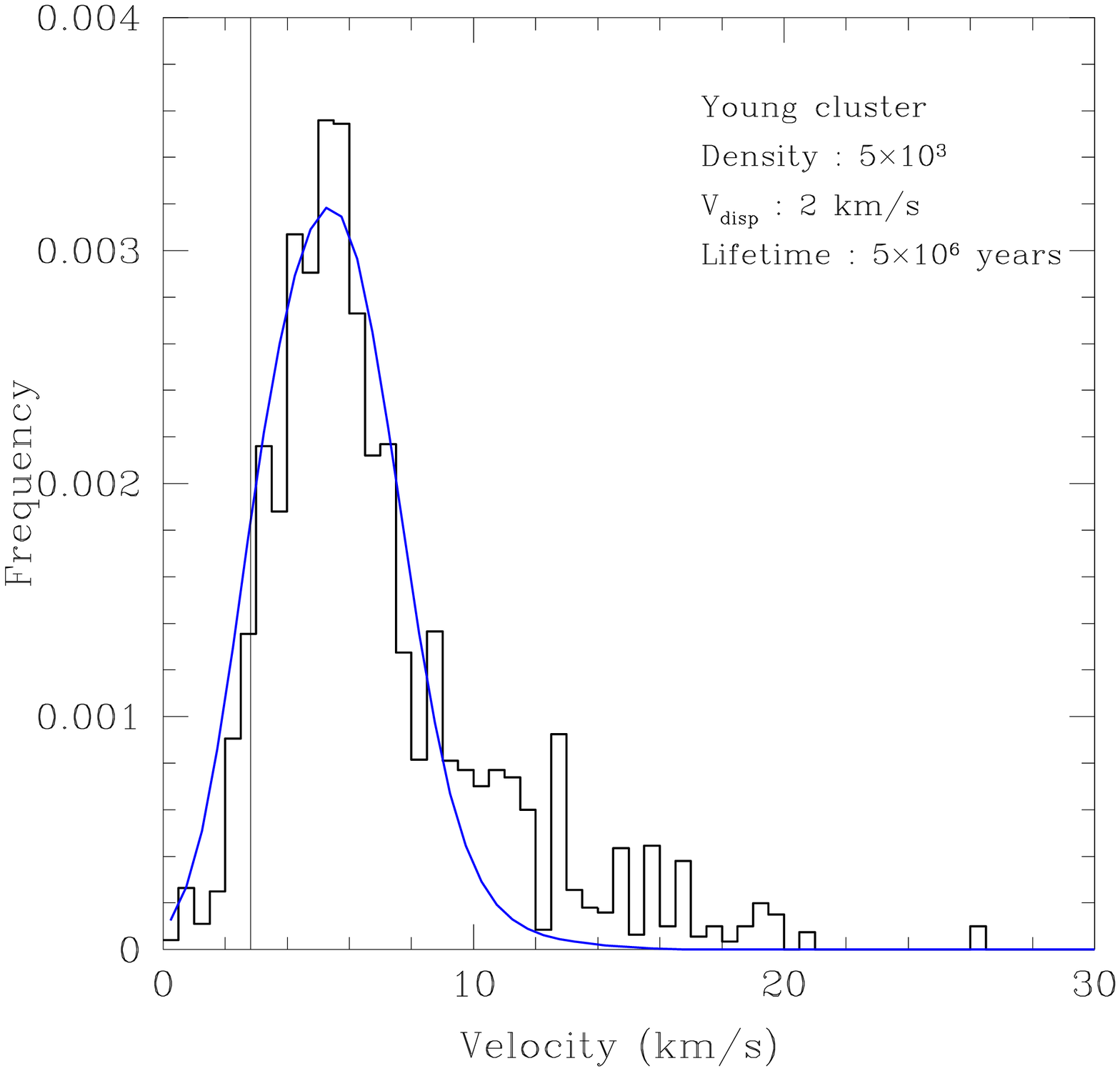,width=3.truein,height=2.3truein}}
}
\caption{\label{vels} The velocity distributions
for the populations of free floating planets in each of the 
three cluster environments. The histogram is the distribution of
total velocities. The vertical line in each case shows the
estimated cluster escape velocity. In each case, a fit to the 
velocity distribution has been made. The Globular cluster case is fitted with
a Gaussian and the other two cases are fitted with a function derived from the 
initial velocity distribution of the ionised systems. See text for details.}
\end{figure}

Graphs of the velocity distributions in various clusters are shown
(Figure~\ref{vels}). The distributions are normalised to the total
number of planets. It is interesting to compare the distributions to
the estimated escape velocity for the cluster (vertical line).  It is
apparent that whilst the globular cluster will retain the bulk of its
free floating planets, most liberated planets in the young cluster or
open cluster will tend to escape within a crossing time. In
Table~\ref{results}, the fraction of planets liberated in each cluster
has been broken down according to whether the planet subsequently
escapes the cluster or not.

We note here that a change in the assumed outer edge of the planetary
orbit distribution, for example truncating the outer orbital radius
closer in, would of course lead to a modification of the final
velocity distribution. The more distant objects are more prone to
disruption, but this is offset by their being less numerous due to the
flat-log initial distribution of orbits.  Truncating the initial
orbits at 20AU rather than 50AU would reduce the fraction of liberated objects to around 
50\% for the globular cluster, 25\% for the open cluster or around 2\% for the
young cluster. This reduction would also tend to affect
the low-velocity population more than the high-velocity tail, since the
high velocity objects come predominantly from the inner orbits.

Also shown in Figure~\ref{vels} are fits to the distributions. The
globular cluster case is reasonably well fitted with a Gaussian. The
other two distributions do not resemble Gaussians, and can only be
poorly represented by a Maxwellian. The distributions shown in these
cases were constructed by taking the product of the initial planetary
orbital velocity distribution (including the stellar velocity
dispersion), and the observed cross section as a function of initial
velocity, and then convolving with a Gaussian. The amplitude of the
distribution and the sigma for the Gaussian were then left as free
parameters in the fit. These distributions do not represent the
observed distribution entirely satisfactorily (they don't reproduce
the high velocity tail), but they serve to illustrate the essential
difference between the high velocity globular cluster case and the low
velocity clusters. In the high velocity dispersion environment of the
globular cluster, the emerging planetary velocity dispersion is
dominated by the stellar scattering, whereas in the open cluster or
young cluster environment the ionised planetary population retains a
memory of the initial Keplerian orbital velocity distribution. It is
this effect which leads to the liberated population escaping from the
low velocity dispersion clusters.

\subsection{The effect of varying planetary masses}
\label{masseffect}

We tested the effects of the restricted three-body assumption for some
specific cases using a three-body Runge-Kutta code and various
planetary masses. It was found that for systems where the planets were
retained by the parent star, the final binding energies differed by at
most a few percent between the massless planet case and the three-body
code with a mass of 0.001M$_{\odot}$ (i.e. 1 Jupiter mass). We also
examined cases where the planetary system was ionised, and
investigated to what extent changing the planetary mass affected the
final escape velocity. The effect was found to be usually modest for
the range of masses applicable to planets (1 to 10 Jupiters), but
could be critical in certain circumstances.  The escape velocity
usually decreased as the planet mass was increased, although there
were cases where the opposite occured. Several cases were found
where modest changes of planet mass produced critical changes in the
escape velocity or changed the encounter outcome from ionised to
retained or exchanged.  These were all distant interactions, in which the
closest approach of the perturbing star to the parent star was greater
than the initial planetary orbit. In these cases, ionisation is of
course sensitive to the encounter conditions, and only a minority of
systems in these encounters were ionised. We therefore conclude that 
the velocity distributions presented would not be changed dramatically 
for any realistic population of planets (up to 10M$_{Jup}$).

\subsection{The bound population: Separation and eccentricity distributions}

\begin{figure*}
\hbox{
\psfig{{figure=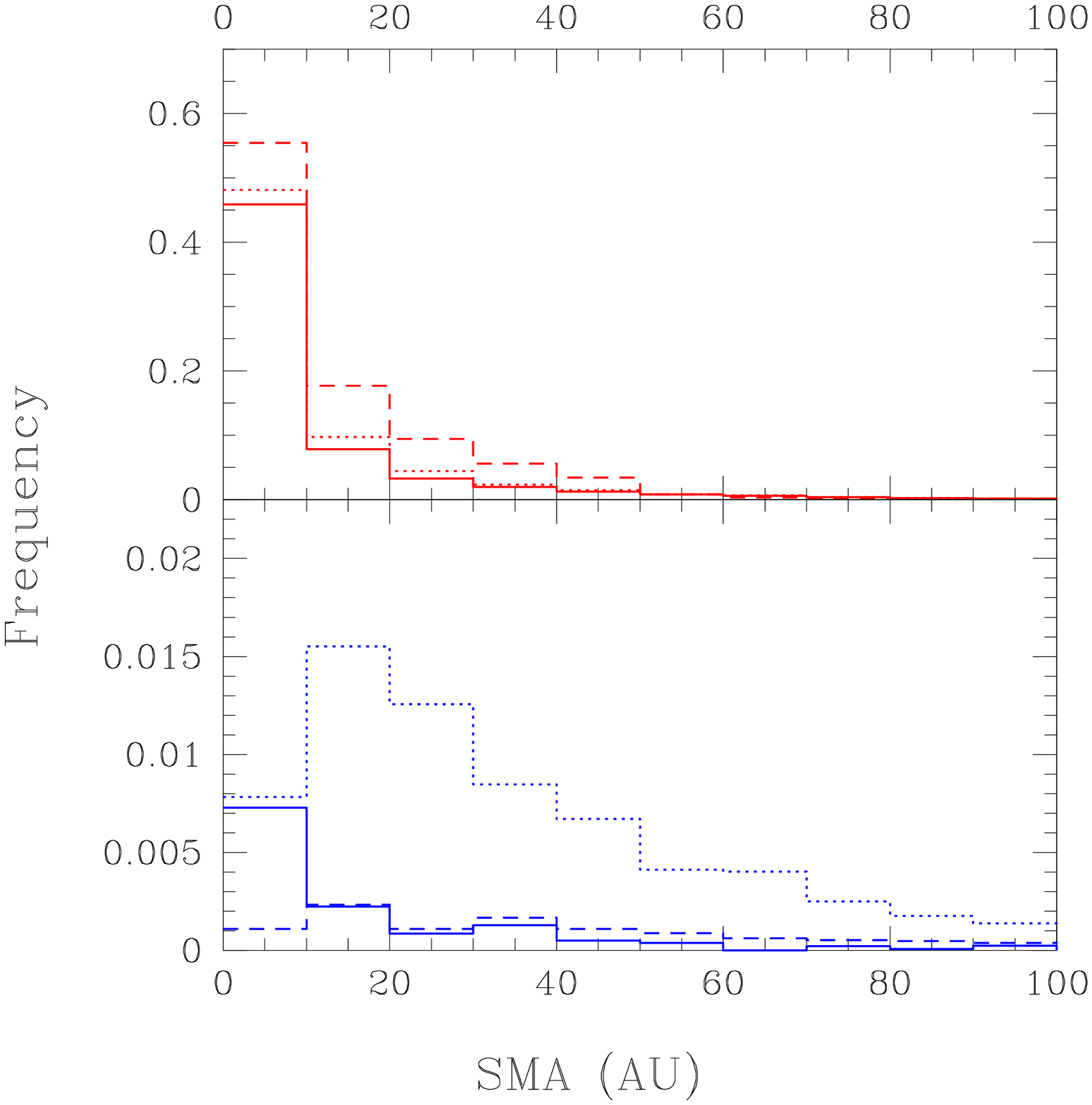,width=3.truein,height=3.truein}}
\psfig{{figure=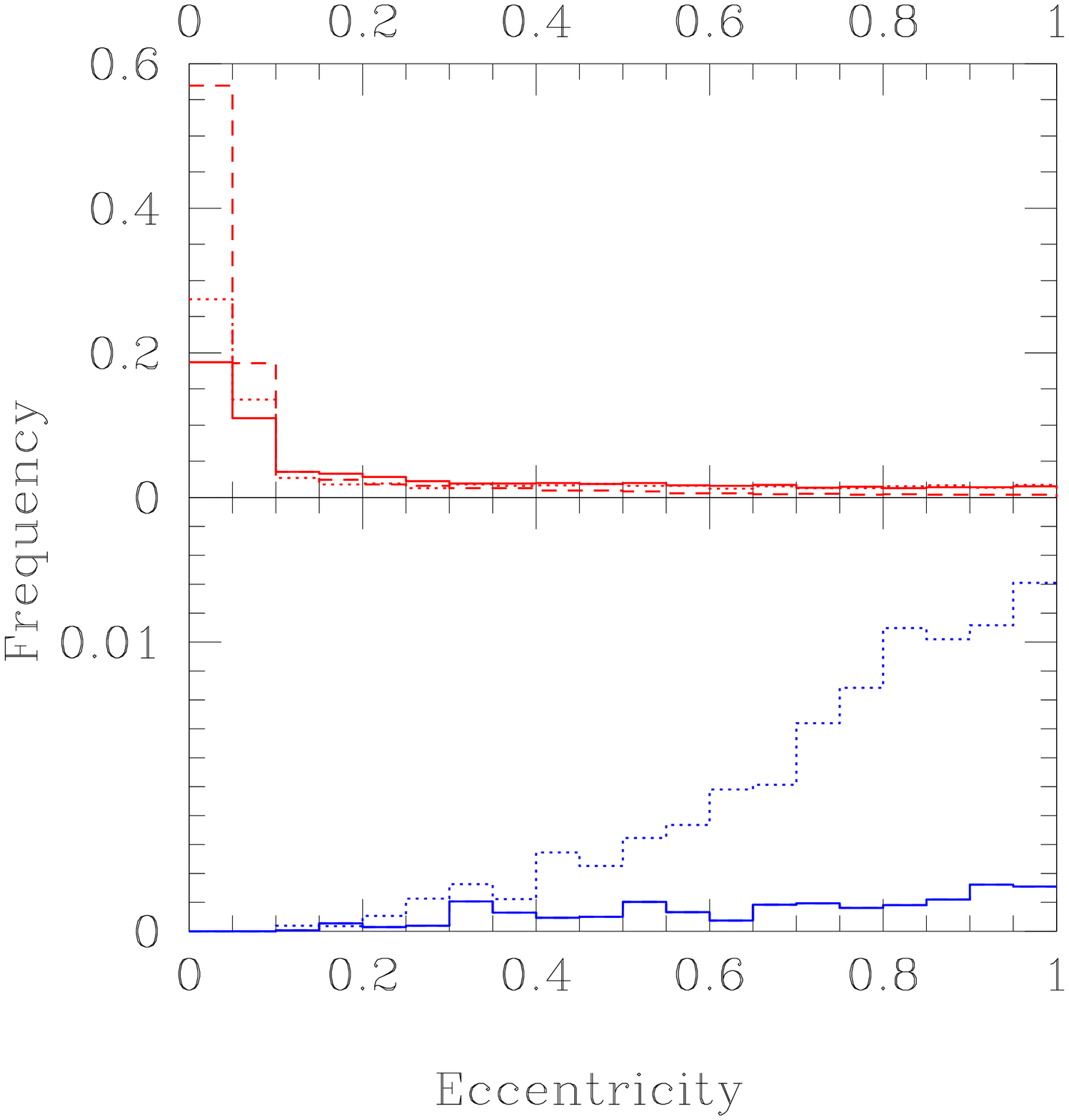,width=3.truein,height=3.truein}}
}
\caption{\label{eccseps} Distributions of semimajor axis (left) and 
eccentricities (right) for surviving
planetary systems. Top panel: systems retained in an encounter, bottom
panel: systems exchanged. The solid line is the globular cluster case,
the dotted line is the open cluster case, and the dashed line is the
young cluster.  The frequency has been normalised to the total number
of systems.  }
\end{figure*}



In Figure~\ref{eccseps} we show the distributions of
separation and eccentricity for the planets which survive
encounters. Separate distributions are shown for the cases where
objects are retained by the parent star and where they are
exchanged. As might be expected, the planets retained by the parent
tend to lie in close orbits. The planets captured by the interloper
occupy a flatter separation distribution. A similar trend is seen in
eccentricity. The retained planets have nearly circular orbits, the
exchanged ones have a flat eccentricity distribution. The highly
eccentric systems and captured systems with large separations will of
course be much more vulnerable to disruption in subsequent encounters.
The effects of scattering on the population of bound planetary
populations in open clusters was investigated in some depth by 
Laughlin \& Adams (1998).

\section{Conclusions}

We have investigated how a population of free-floating planets can be
generated by stellar encounters in different cluster environments. We
have found that in globular clusters a relatively high fraction of any
planetary population is likely to be liberated by encounters over the
cluster lifetime, and furthermore that the majority of these systems
should be retained in the cluster at least until they are lost through
two body relaxation after several thousand crossing times.

In the less dense environments of an open cluster or young star
forming cluster, planet liberation was found to be less efficient,
although still capable of producing a significant population of free
floating planets. However, it was found that these objects were
liberated at too high a velocity to remain bound in the cluster.  In
each case, only a fraction of a percent of the planetary population
was liberated but remained bound to the cluster. This suggests that
there should not be substantial numbers of free floating planets in
such environments. Furthermore, any such objects which were observed 
in stellar clusters would be expected to have a higher velocity than the
cluster stars, and so to be found predominantly in the outer regions
far from the cluster core.

This has a bearing on the recent discovery of substellar
objects in $\sigma$ Orionis (Zapatero-Osorio et al, 2000). The objects
found in this study were typically many Jupiter masses, although
some were as little as 5M$_{Jup}$. It is not
clear whether such massive objects should better be regarded as
planets or as brown dwarfs. Our results imply that they have probably formed 
independently rather than in a
protostellar disc.  The higher mass of some of the $\sigma$ Orionis objects
(up to 50M$_{Jup}$) should not strongly affect the escape velocities
except in the a few cases of distant encounters (See
Section~\ref{masseffect}).

We note finally that the objects escaping from stellar clusters will
form a population of fast moving unbound planets in the Galactic disc.
However, this would not be expected to form a significant contribution to 
the total mass of the Galaxy.

\section{Acknowledgements}

We thank the Asgard staff at ETH for providing the computer 
resources used in this work. We also thank Peter Messmer and Simon Hall
for useful discussions, and the referee, Derek Richardson,
for his rapid and helpful report.

\label{lastpage}

\end{document}